\documentclass[twocolumn,aps,prl,superscriptaddress,showpacs,tightenlines]{revtex4}
\usepackage{amsmath}
\usepackage{amsfonts}
\usepackage{graphicx}
\usepackage{epsfig}
\usepackage{color}
\usepackage[colorlinks,citecolor=blue]{hyperref}

\begin{document}

\title{Squeezed Optomechanics with Phase-matched Amplification and Dissipation}

\author{Xin-You L\"{u}}
\email{xinyoulu@gmail.com}
\affiliation{School of physics, Huazhong University of Science and Technology, Wuhan 430074, China}
\affiliation{iTHES, RIKEN, Saitama 351-0198, Japan}
\author{Ying Wu}
\email{yingwu2@126.com}
\affiliation{School of physics, Huazhong University of Science and Technology, Wuhan 430074, China}
\author{J.R. Johansson}
\affiliation{iTHES, RIKEN, Saitama 351-0198, Japan}
\author{Hui Jing}
\affiliation{CEMS, RIKEN, Saitama 351-0198, Japan}
\affiliation{Department of Physics, Henan Normal University, Xinxiang 453007, China}
\author{Jing Zhang}
\affiliation{CEMS, RIKEN, Saitama 351-0198, Japan}
\affiliation{Tsinghua National Laboratory for Information Science and Technology, Beijing 100084, China}
\author{Franco Nori}
\affiliation{CEMS, RIKEN, Saitama 351-0198, Japan}
\affiliation{Department of Physics, The University of Michigan, Ann Arbor, Michigan 48109-1040, USA}

\begin{abstract} 
We investigate the nonlinear interaction between a squeezed cavity mode and a mechanical mode in an optomechanical system (OMS) that allows us to selectively obtain either a radiation-pressure coupling or a parametric-amplification process. The squeezing of the cavity mode can enhance the interaction strength into the single-photon strong-coupling regime, even when the OMS is originally in the weak-coupling regime. Moreover, the noise of the squeezed mode can be suppressed completely by introducing a broadband-squeezed vacuum environment that is phase-matched with the parametric amplification that squeezes the cavity mode. This proposal offers an alternative approach to control OMS using a squeezed cavity mode, which should allow single-photon quantum processes to be implemented with currently available optomechanical technology. Potential applications range from engineering single-photon sources to nonclassical phonon states.
\end{abstract}
\maketitle

Cavity optomechanics has progressed enormously in recent years~\cite{reviews}, with achievements including cooling of mechanical modes to their quantum ground states~\cite{coolingExp1,coolingExp2}, demonstration of optomechanically-induced transparency~\cite{OITExp1, OITExp2}, coherent state transfer between cavity and mechanical modes~\cite{stateconversion1, stateconversion2, stateconversion3, stateconversion4}, and the realization of squeezed light~\cite{squeezingExp1, squeezingExp2,squeezingExp3}. In these experiments, a strong linearized optomechanical coupling is obtained under the condition of strong optical driving. However, the intrinsic nonlinearity of the radiation-pressure coupling in these OMSs is negligible~\cite{Huang2009,Xiong2012,Wang2012, Tian2012, Ma2014, Jiang2014, Jing2014}.

To explore the intrinsic nonlinearity of the optomechanical interaction, much theoretical research has recently focused on the single-photon strong-coupling regime, where the single-photon optomechanical-coupling strength $g_0$ exceeds the cavity decay rate $\kappa$. In this regime, several interesting single-photon quantum processes are predicted, for both the optical and the mechanical modes. For example: photon blockade, the preparation of the nonclassical states of the optical and mechanical modes, multi-phonon sidebands, and quantum state reconstruction of the mechanical oscillator~\cite{Bose1997,Bouwmeester2003,XWXu2013,Rabl2011,Girvin2011,Jieqiao2012,Ludwig2012,Jieqiao2013,Yuxi2013,Marquardt2013,XYL2013,Rabl2012,Lukin2013,Clerk2013,Jieqiao2014}. However, these effects have not yet been realized experimentally due to the intrinsically weak radiation-pressure coupling in current OMSs, i.e., $g_0\ll \kappa$. To achieve $g_0\sim\kappa$, it has been proposed to use the collective mechanical modes in transmissive scatter arrays~\cite{Xuereb2012,Xuereb2013}. The ratio $g_0/\kappa$ may also be increased in superconducting circuits using the Josephson effect, but such devices are limited to electromechanical systems~\cite{Sillanp2014,Johansson2014,Nation2014}. Moreover, postselected weak measurements \cite{Gang2014} and optical coalescence \cite{Genes} could also be used to increase the effective linear and quadratic optomechanical interactions, respectively.

Here we present a method to reach the single-photon strong-coupling regime in an OMS, which is originally in the weak-coupling regime. In contrast to normal optomechanics, we focus on the nonlinear interaction between a parametric-amplification-squeezed cavity mode and a mechanical mode. We obtain an optomechanical coupling that selectively can take the forms of radiation-pressure or a parametric-amplification process. Physically, a single-photon state in the squeezed cavity mode corresponds to an exponentially-growing number of photons in the original cavity mode, as a function of increasing squeezing strength. Consequently, the optomechanical interaction in units of the squeezed-cavity-mode photons can be enhanced, e.g., into the single-photon strong-coupling regime, by tuning the intensity (or frequency) of the driving field that induce the squeezing. 

In addition, we show that the noise of the squeezed cavity mode can be suppressed by introducing a broadband-squeezed vacuum~\cite{Schnabel2013,Devoret2009} with a reference phase matching the phase of the driving field. Under these conditions of enhanced coupling strength and suppressed noise, it should be feasible to implement single-photon quantum processes even in an originally weakly-coupled OMS. Our proposal is also suitable for electromechanical systems with squeezed-vacuum reservoirs for superconducting resonators~\cite{Siddiqi2013}. Note that a broadband-squeezed vacuum also can suppress the radiative decay of atoms~\cite{Gardiner1986,Georgiades1995,Dayan2004} or the artificial atoms~\cite{Siddiqi2013}, and can be used to squeeze the mechanical modes in OMSs~\cite{Zoller1999,GXLi2013}.

\begin{figure}
\includegraphics[width=7.5cm]{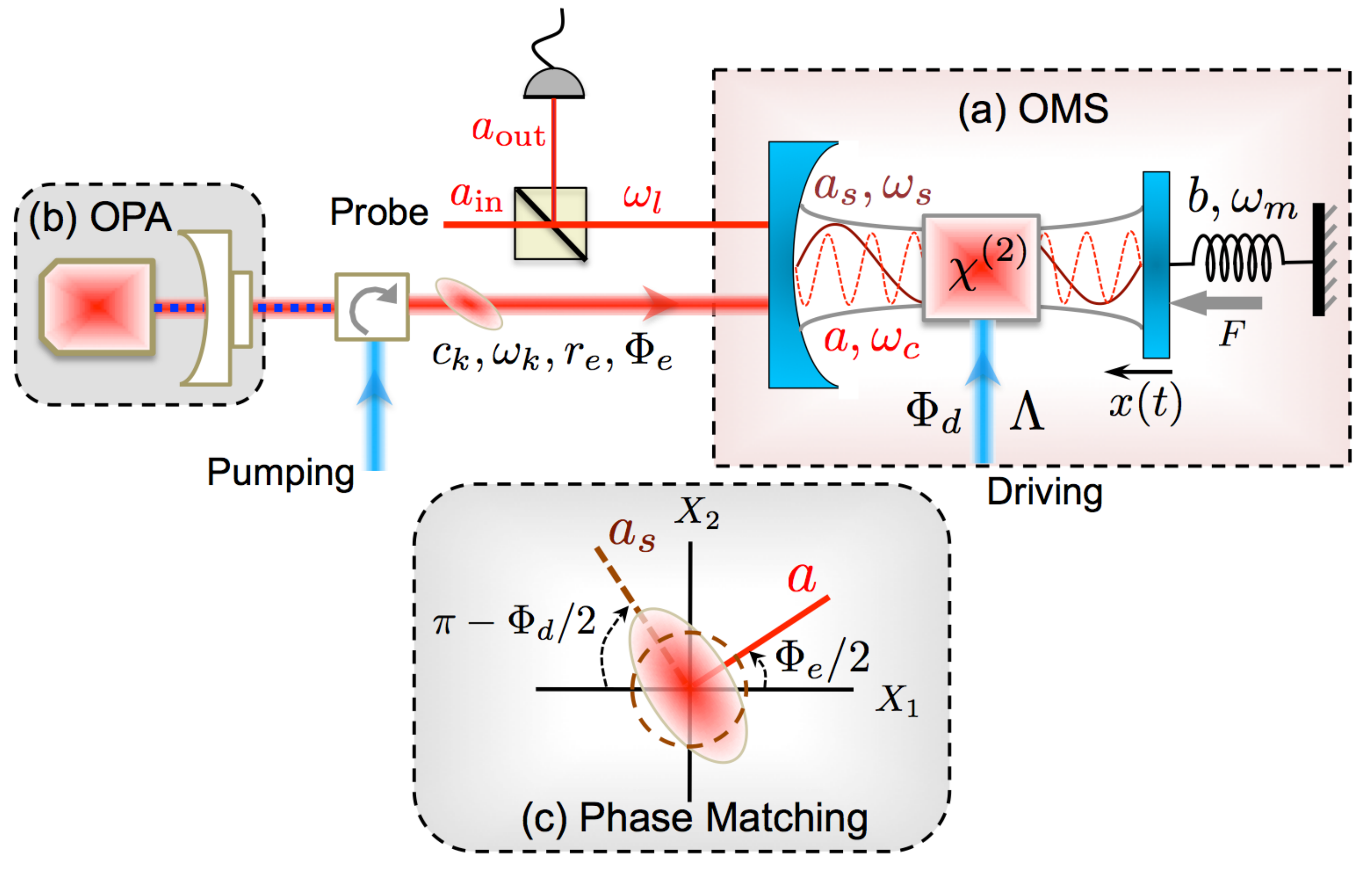}
\caption{(Color online) (a) A schematic illustration of an OMS with mechanical mode $b$ (driven by a force $F$), main cavity $a$ and a squeezed cavity-mode $a_s$ induced by driving a $\chi^{(2)}$ nonlinear medium with frequency $\omega_d$, amplitude $\Lambda$, and phase $\Phi_d$. Here $a_{\rm in}$ and $a_{\rm out}$ are the input and output of a weak probe field with frequency $\omega_l$. (b) A broadband squeezed-vacuum-field $c_k$ with frequency $\omega_k$ (generated by an OPA) interacts with $a$. The squeezing parameter and reference phase are $r_e$ and $\Phi_e$. (c) The phase-matching condition $\Phi_e-\Phi_d=\pm n\pi$ $(n=1, 3, 5, \dots)$ for suppressing the noise of $a_s$, is indicated by the squeezing directions.}
\label{fig1}
\end{figure}

\emph{System.---} We consider an OMS depicted in Fig.\,\ref{fig1}(a) with the Hamiltonian ($\hbar=1$)
\begin{equation}
H=H_{c}+H_{m}-g_{0}a^{\dagger}a(b^{\dagger}+b),\label{H_original}
\end{equation}
where $a$ ($a^{\dagger}$) and $b$ ($b^{\dagger}$) are the annihilation (creation) operators of the cavity mode and the mechanical mode, respectively. The optical cavity (with resonance frequency $\omega_{c}$) contains a $\chi^{(2)}$ nonlinear medium that is pumped with driving frequency $\omega_d$, amplitude $\Lambda$, and phase $\Phi_d$. Its Hamiltonian can be written as $H_{c}=\Delta_{c}a^{\dagger}a +\Lambda(a^{\dagger2}e^{-i\Phi_d}+a^2e^{i\Phi_d})$, with $\Delta_{c}=\omega_c-\omega_d/2$ in a frame rotating with $\omega_d/2$. The Hamiltonian of the mechanical mode $H_{m}=\omega_{m}b^{\dagger}b+F(b^{\dagger}+b)$ (with mechanical frequency $\omega_{m}$) contains a constant force $F$ that cancels a force induced by the parametric amplification (see below). The third term in Eq.~(\ref{H_original}) describes the radiation-pressure interaction between the cavity and the mechanical modes with coupling strength $g_{0}$~\cite{CKLawPRA1995}. Here the $\chi^{(2)}$ nonlinearity is used to induce a squeezed cavity mode. It could also be used to enhance optomechanical cooling~\cite{Agarwal2009}, induce genuine tripartite entanglement~\cite{Paternostro}, or impact the classical dynamics of OMSs~\cite{Benlloch}.

As shown in Fig.\,\ref{fig1}(b), an optical parametric amplification (OPA) is introduced to generate a broadband-squeezed vacuum field $c_k$ (with central frequency $\omega_c$), which is injected into the cavity. Here $r_e$ and $\Phi_e$ are the squeezing parameter and reference phase of this squeezed environment, respectively, corresponding to the intensity and phase of the pump field. Experimentally, optical (microwave) light with squeezing bandwidth up to GHz~\cite{Schnabel2013} (tens of MHz~\cite{Siddiqi2013}) has been realized. This is much larger than the typical linewidth of optical (microwave) cavities, i.e., MHz (hundreds of kHz). From the point of view of the cavity, the squeezed input field is well approximated as having infinite bandwidth~\cite{Siddiqi2013}. The dissipation caused by the system-bath coupling can then be described by the Lindblad superoperators $\kappa(N+1)\mathcal{D}[a]\rho+\kappa N\mathcal{D}
[a^{\dagger}]\rho-\kappa M\mathcal{G}[a]\rho-\kappa M^{*}\mathcal{G}[a^{\dagger}]\rho$ (cavity damping) and $\gamma(\bar{n}^{m}_{\rm th}+1)\mathcal{D}[b]\rho+\gamma\bar{n}^{m}_{\rm th}\mathcal{D}[b^{\dagger}]\rho$ (mechanical damping) in the master equation. Here $\mathcal{D}[o]\rho=o\rho o^{\dagger}-(o^{\dagger}o\rho+\rho o^{\dagger}o)/2$, $\mathcal{G}[o]\rho=o\rho o-(oo\rho+\rho oo)/2$, $\kappa$ and $\gamma$ are the cavity and mechanical decay rates, respectively, and $\bar{n}^m_{\textrm{th}}$ is the thermal phonon number of the mechanical mode. The mean photon number of the broadband squeezed field is $N=\sinh^2(r_e)$, and $M=\sinh(r_e)\cosh(r_e)e^{i\Phi_e}$ describes the strength of the two-photon correlation \cite{book2002}.

\emph{Parametric-amplification-induced strong optomechanical coupling.---}
Parametric amplification in the cavity introduces a preferred squeezed cavity mode $a_s$ that satisfies a squeezing transformation $a=\cosh(r_d)a_s-e^{-i\Phi_d}\sinh(r_d)a_s^{\dagger}$, with $r_d=(1/4)\ln[(\Delta_c+2\Lambda)/(\Delta_c-2\Lambda)]$. In terms of $a_s$, Hamiltonian (\ref{H_original}) can be rewritten as
\begin{equation}
\!\!\!\!H\!\!\!=\!\omega_sa^{\dagger}_sa_s\!+\omega_mb^{\dagger}b\!-g_sa^{\dagger}_sa_s(b^{\dagger}+b)\!+\frac{g_p}{2}(a^{\dagger2}_s+a^2_s)(b^{\dagger}+b),\label{H_squeezed}
\end{equation}  
where the cavity Hamiltonian $H_c$ has been diagonalized by the squeezing transformation, and is expressed as an oscillator with a controllable frequency 
$\omega_s=(\Delta_c-2\Lambda)\exp(2r_d)$.
Here we have chosen $F=g_0{\sinh^2(r_d)}$ to cancel an induced force applied to the mechanical oscillator.
The third and fourth terms in Eq.\,(\ref{H_squeezed}) describe the standard optomechanical radiation-pressure and parametric-amplification interactions, respectively, with the controllable strengths 
\begin{subequations}\label{opmc}
\begin{align}
g_s&=\frac{g_0\Delta_c}{\sqrt{\Delta^2_c-4\Lambda^2}}=g_0\cosh(2r_d),
\\
g_p&=\frac{2g_0\Lambda}{\sqrt{\Delta^2_c-4\Lambda^2}}=g_0\sinh(2r_d).
\end{align}
\end{subequations}
This provides an optomechanical interaction that can be tuned by adjusting the system parameters, such as the frequency detuning $\Delta_c$ and the driving strength $\Lambda$. [The small optical linewidth should be included in Eqs.\,(\ref{opmc}) in the extremely narrow critical regime where $|\Delta_c|$ infinitely approaches 2$\Lambda$. The detailed discussion is omitted because it does not limit the efficiency of our mechanism significantly in practice.]

\begin{figure}
\includegraphics[width=4cm]{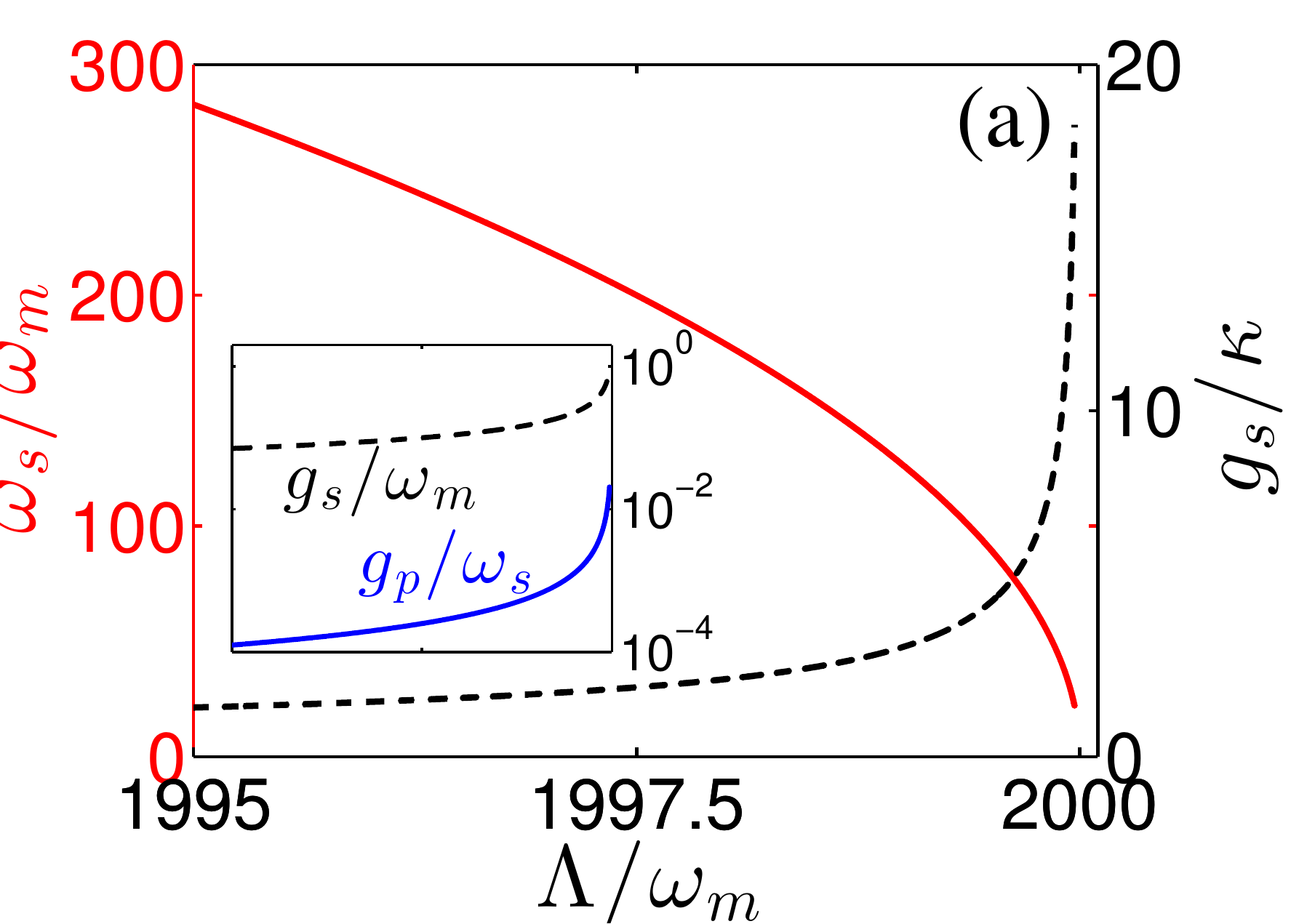}\includegraphics[width=4cm]{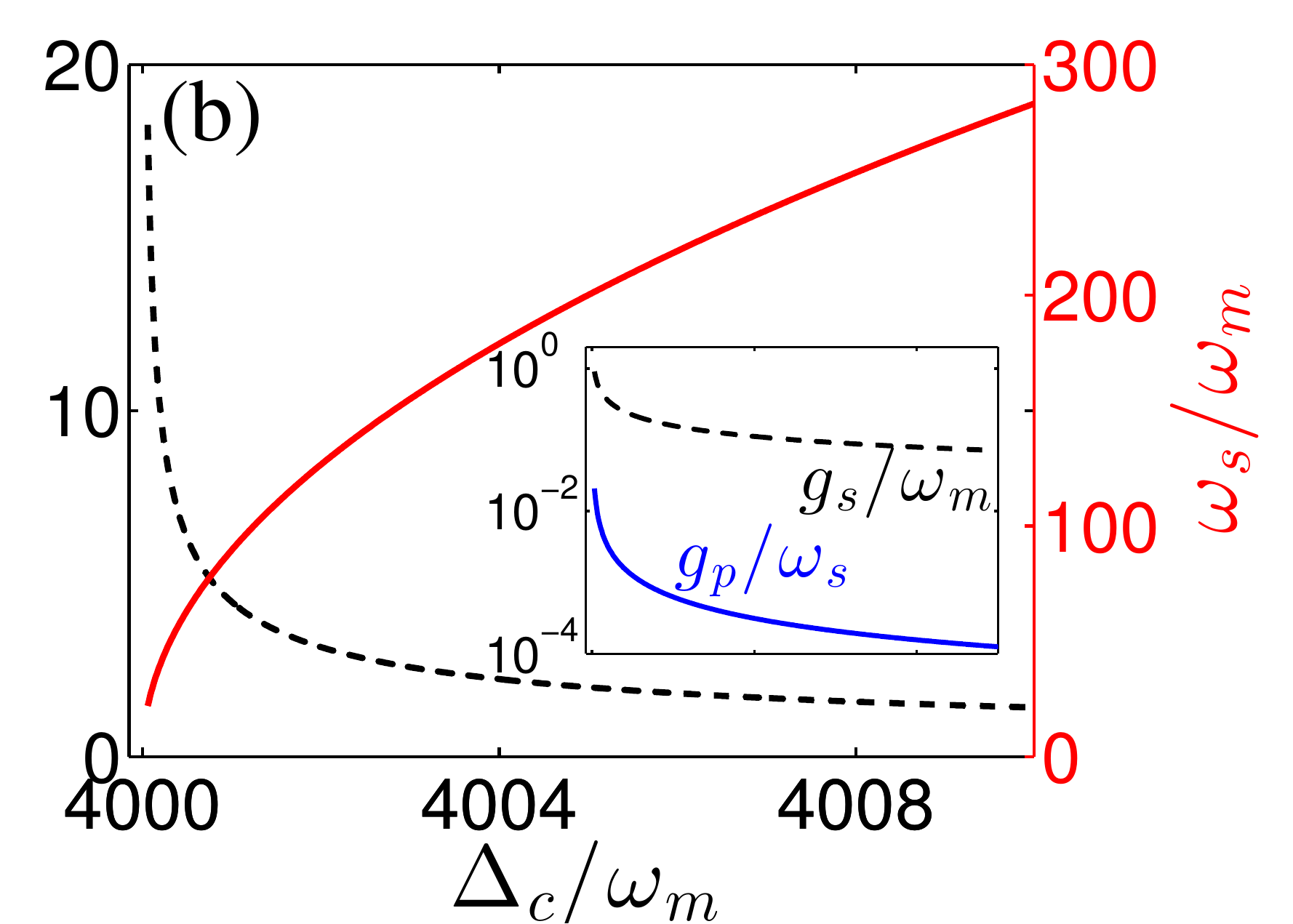}
\includegraphics[width=4cm]{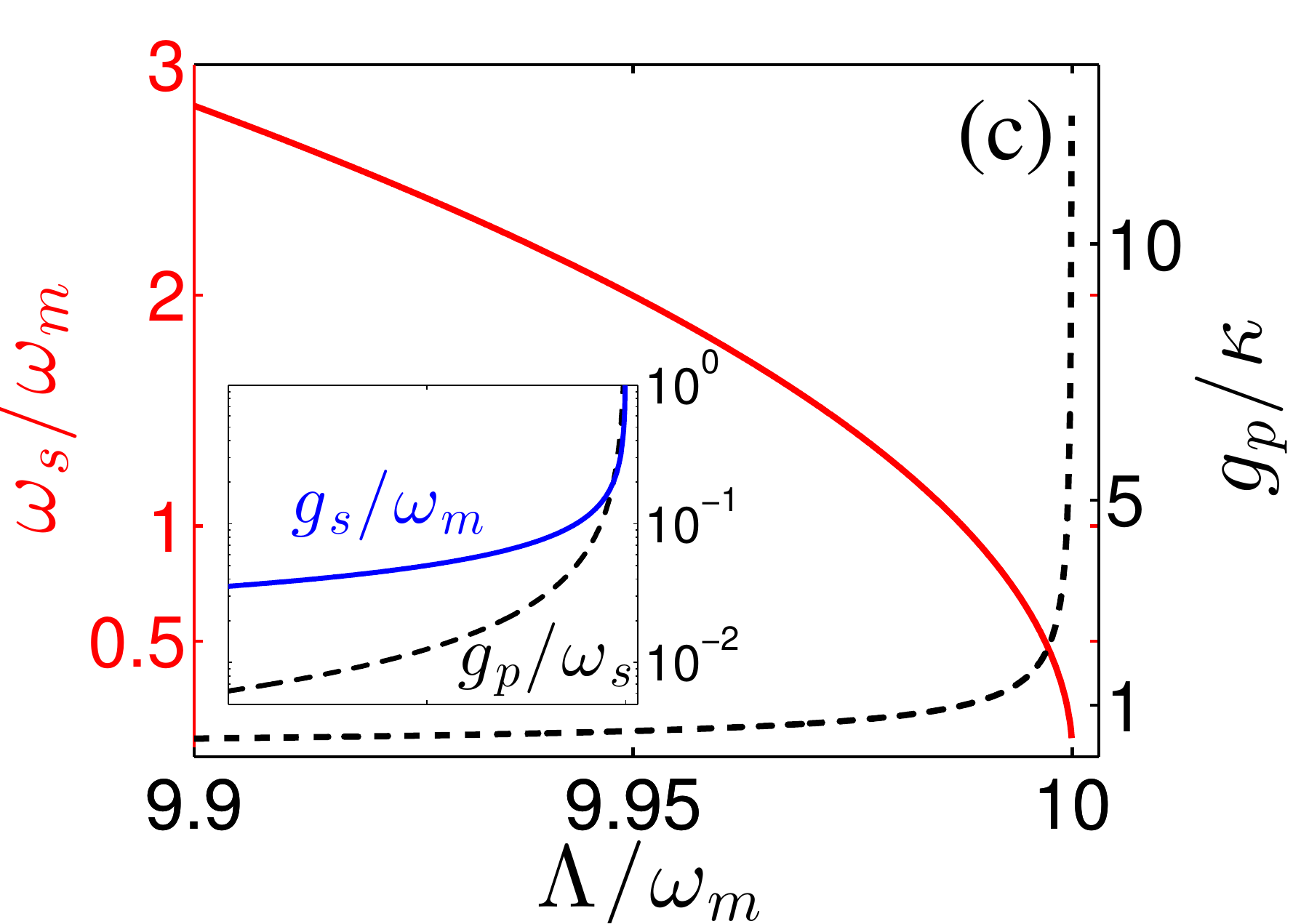}\includegraphics[width=4cm]{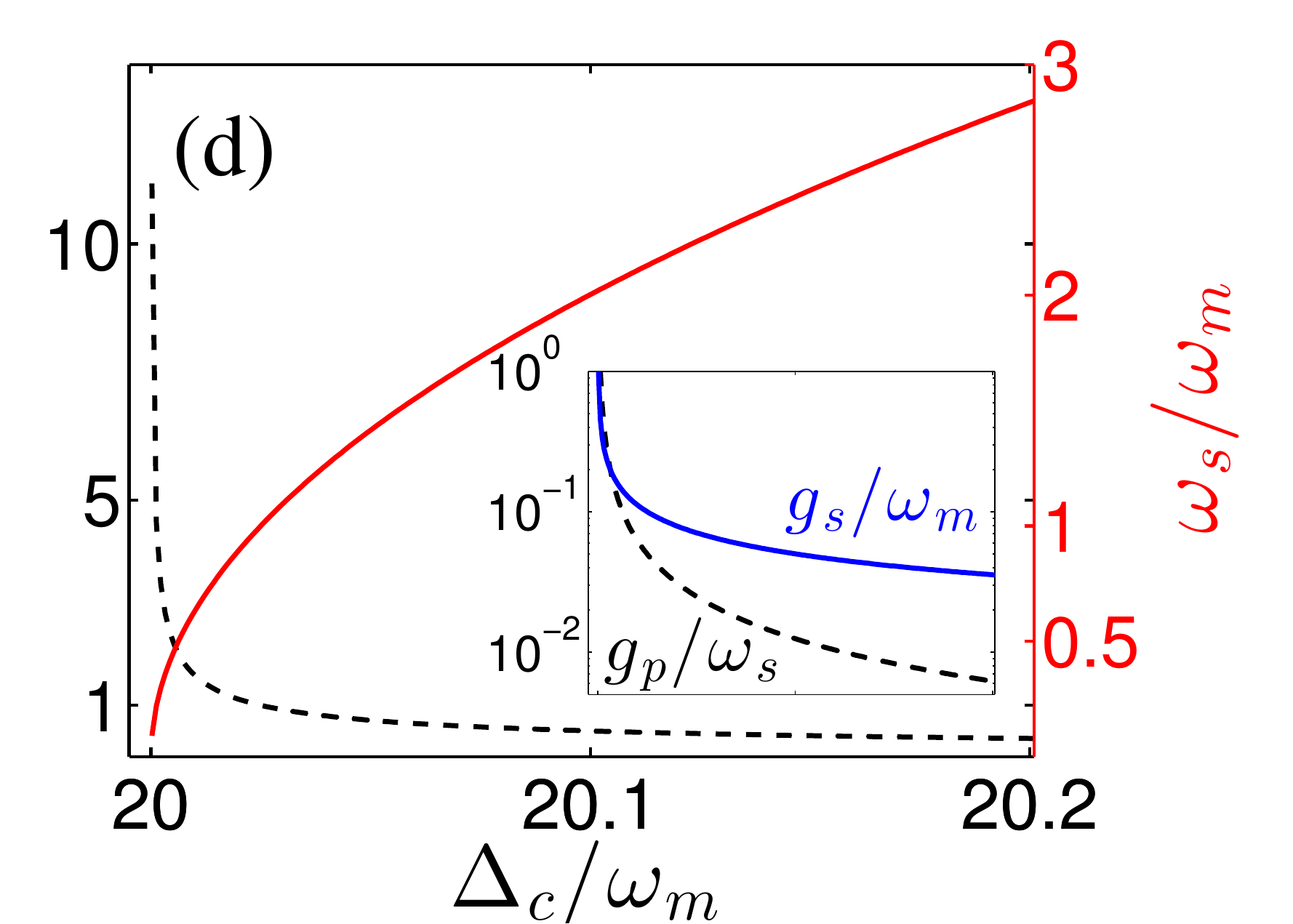}
\caption{(Color online) The optomechanical coupling strengths $g_s/\kappa$, $g_p/\kappa$, and the cavity frequency $\omega_s/\omega_m$ versus driving strength $\Lambda$ and detuning $\Delta_c$. The values $g_s/\omega_m$ and $g_p/\omega_s$ are presented in the inserts. The parameters are $g_0=0.005\omega_m$, $\kappa=0.05\omega_m$, $\gamma=10^{-4}\omega_m$, and (a) $\Delta_c=4000\omega_m$, (c) $\Delta_c=20\omega_m$, (b) $\Lambda=2000\omega_m$, (d) $\Lambda=10\omega_m$.}
\label{fig2}
\end{figure}

On one hand, the parametric interaction [last term of Eq.\,(\ref{H_squeezed})] can be suppressed by adjusting $\Delta_c$ or $\Lambda$ so that $\omega_s\gg g_p,\omega_m$. Under a rotating-wave approximation (RWA), we obtain a standard optomechanical Hamiltonian
\begin{align}
H_{\rm OMS}=\omega_sa^{\dagger}_sa_s+\omega_mb^{\dagger}b-g_sa^{\dagger}_sa_s(b^{\dagger}+b),\label{H_OMS}
\end{align}
by safely neglecting the terms that oscillate with high frequencies, $2\omega_s\pm\omega_m$. In this case, the single-photon optomechanical-coupling strength $g_s$ could be significantly enhanced  (approximately three orders of magnitude) and reach the strong-coupling regime, i.e., $g_s>\kappa$ [see Figs.\,\ref{fig2}(a,b)]. This enhancement is due to that a single-photon state in the squeezed mode $|1\rangle_s$ corresponds to an exponentially-growing number of photons in the original cavity, as a function of increasing squeezing strength, i.e., $_s\langle1|a^{\dagger}a|1\rangle_s\rightarrow \cosh(2r_d)$. The radiation pressure of a single squeezed photon on the mechanical resonator is therefore correspondingly increased, which effectively enhances the optomechanical coupling between the mechanical mode and the {\em squeezed} cavity mode.

On the other hand, we could also suppress the radiation-pressure interaction by adjusting $\Delta_c$ or $\Lambda$, so that $g_s/\omega_m,g_p/\omega_s\ll 1$ and $\omega_s\approx \omega_m/2$ [see Figs.\,\ref{fig2}(c,d)]. Under a RWA, Hamiltonian (\ref{H_squeezed}) is simplified to a resonant photon-phonon parametric interaction, i.e., $H_{\rm PI}=\omega_sa_s^{\dagger}a_s+\omega_mb^{\dagger}b+g_p(a^2_sb^{\dagger}+ba^{\dagger2}_s)$, in the strong-coupling regime $g_p>\kappa$. This could potentially be used for highly efficient down-conversion of a single phonon into an entangled photon pair.

\emph{Suppressing the cavity noise with phase matching.---}
Expressing the system-bath interaction in terms of $a_s$, the system master equation can be rewritten as
\begin{eqnarray}
\dot{\rho}\!=\!\!\!&-&\!\!i[H,\rho]\!+\!\kappa(N_s\!+\!1)\mathcal{D}[a_s]\rho\!+\!\kappa N_s\mathcal{D}[a_s^{\dagger}]\rho\!-\!\kappa M_s\mathcal{G}[a_s]\rho
\nonumber \\
&-&\!\!\kappa M_s^{*}\mathcal{G}[a_s^{\dagger}]\rho+\gamma\bar{n}^m_{\textrm{th}}\mathcal{D}[b^{\dagger}]\rho+\gamma(\bar{n}^m_{\textrm{th}}+1)\mathcal{D}[b]\rho,\label{rho}
\end{eqnarray}
where $H$ is given by Eq.\,(\ref{H_squeezed}). Here $N_s$ and $M_s$ denote the effective thermal noise and two-photon-correlation strength, respectively, given by (setting $\Phi=\Phi_e-\Phi_d$)
\begin{subequations}
\begin{align}
N_{s}=&\sinh^2(r_d)\cosh^2(r_e)+\cosh^2(r_d)\sinh^2(r_e)\nonumber
\\
&+\frac{1}{2}\cos(\Phi)\sinh(2r_d)\sinh(2r_e),\label{nth}
\\
M_{s}=&e^{i\Phi_d}\left[\cosh(r_d)\cosh(r_e)+e^{-i\Phi}\sinh(r_d)\sinh(r_e)\right]\nonumber
\\
&\left[\sinh(r_d)\cosh(r_e)+e^{i\Phi}\cosh(r_d)\sinh(r_e)\right].\label{nrot}
\end{align}
\end{subequations}
When $r_d=r_e=r$, $N_{s}$ and $M_s$ simplify to $N_{s}=\sinh^2(2r)[1+\cos(\Phi)]/2$ and $M_s=\exp(i\Phi_d)\sinh(2r)[1+\exp(i\Phi)][\cosh^2(r)+\exp(-i\Phi)\sinh^2(r)]/2$, respectively. This shows that the thermal noise and the two-photon correlation can be suppressed completely (i.e., $N_s, M_s=0$) when $r_d=r_e$ and $\Phi=\pm n\pi$ ($n=1, 3, 5, \dots$). This result can be understood from the phase matching in Fig.~\ref{fig1}(c). The reservoir of the original cavity is squeezed along the axis with angle $\Phi_e/2$, with a squeezing parameter $r_e$. In the basis of the squeezed cavity modes $a_s$, this effect is cancelled by the squeezing (along axis $\Phi_d/2$) induced by the parametric amplification of $a$, when $\Phi_e-\Phi_d=\pm n\pi$ and $r_e=r_d$. That is, the squeezed-vacuum reservoir (ellipse) of $a$ corresponds to an effective vacuum reservoir (circle) of $a_s$. 
 
\begin{figure}
\includegraphics[width=4.3cm]{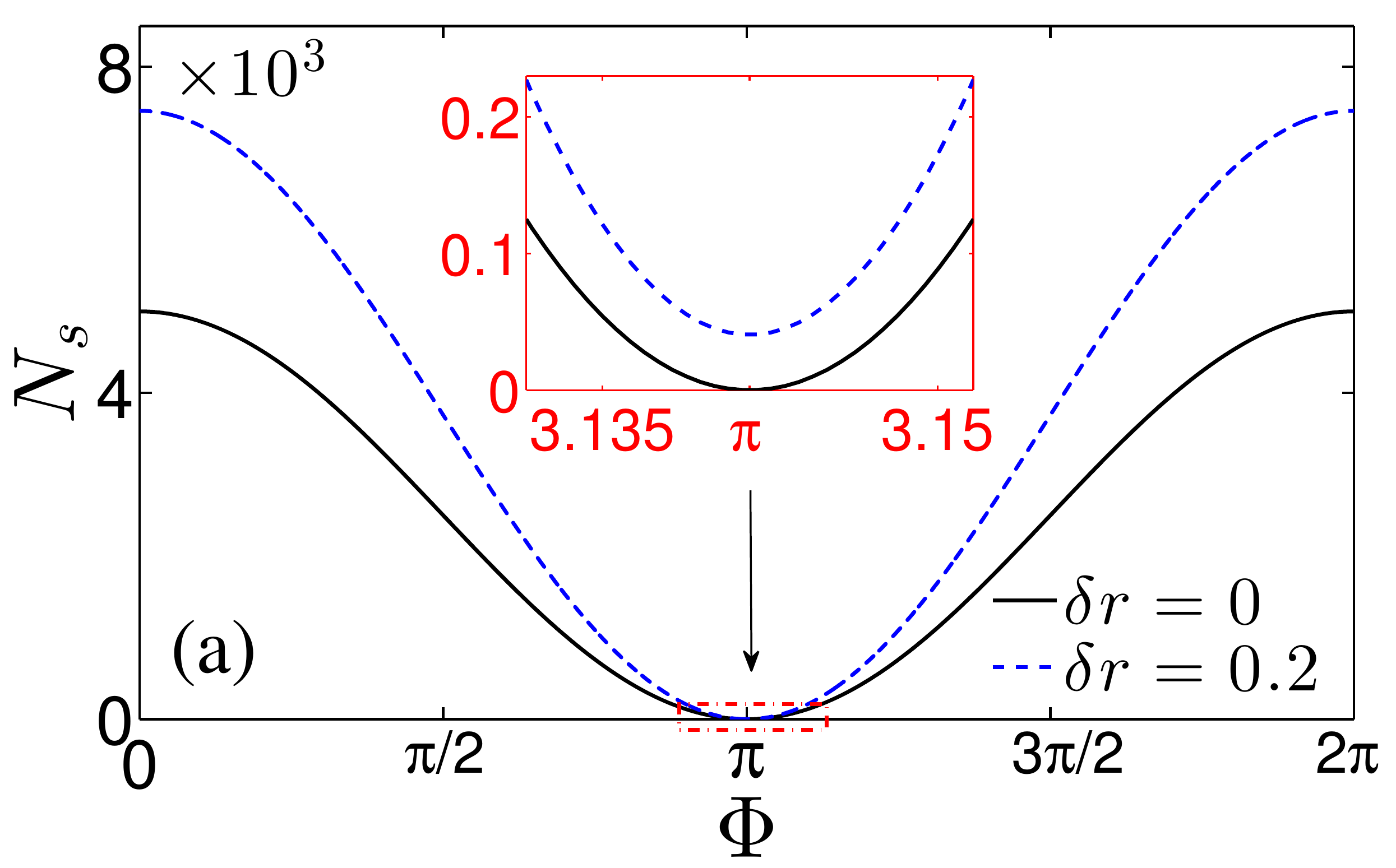}\includegraphics[width=4.3cm]{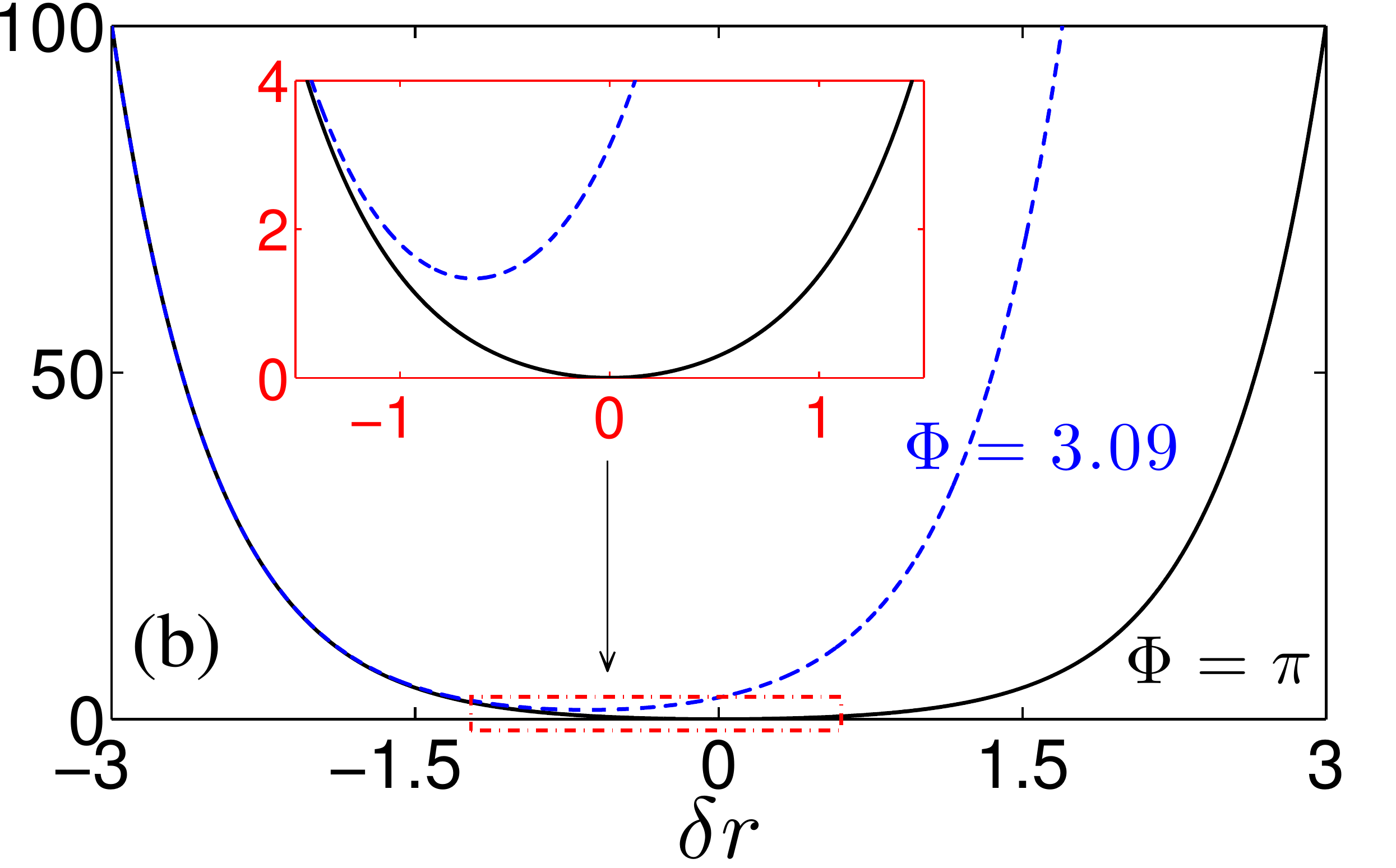}
\caption{(Color online) The effective thermal noise $N_{s}$ versus (a) $\Phi$, (b) $\delta r=r_e-r_d$ for different (a) $\delta r$ and (b) $\Phi$. The insert corresponds to the vicinity of the ideal parameter regime.}
\label{fig3}
\end{figure}

In Fig.~\ref{fig3}, we plot $N_s$ as a function of the phase $\Phi$ and squeezing imbalance $\delta r=r_e-r_d$. Note that the amplitude of $M_s$ has almost the same behavior as $N_s$, and is not plotted here. Fig.~\ref{fig3} shows that the ideal parameters are $\Phi=\pm n\pi$ and $\delta r=0$, which is consistent with our qualitative discussion. Deviating from these ideal parameters, $N_{s}$ increases periodically (exponentially) with increasing $\Phi$ ($\delta r$). The insert of Fig.~\ref{fig3}(b) also shows that the optimal point of $\delta r$ shifts with changing $\Phi$, which can be understood from the third term in Eq.~(\ref{nth}).

\emph{Applications.---} 
To probe the radiation-pressure coupling, one can drive the original cavity mode using a weak probe field with frequency $\omega_l$, amplitude $\epsilon_l$, 
($\epsilon_l\ll\kappa$). The Hamiltonian is $H_{p}=a^{\dagger}e^{-i\omega^s_l t}+ae^{i\omega^s_l t}$ in the frame rotating with $\omega_d/2$, and $\omega^s_l=\omega_l-\omega_d/2$ is the effective frequency of the probe field. Note that, only the squeezed mode $a_s$ is excited when $\omega^s_l\approx\omega_s$, and this is achieved by a joint effect of the probe and driving fields. In this case, the optomechanical coupling strength could be inferred by measuring the steady-state excitation spectrum, i.e., $S(\Delta_s)=({\rm Lim}_{t\rightarrow\infty}\langle a_s^{\dagger}a_s\rangle(t)-N_s)/n_0$ ($n_0=4\epsilon_l^2/\kappa^2$, $\Delta_s=\omega_s-\omega^s_l$) \cite{Rabl2011,Girvin2011,Jieqiao2012}, which has been shifted by a constant $N_s$ when $\Phi\neq\pi$ ($N_s=0$ when $\Phi=\pi$). 

\begin{figure}
\includegraphics[width=8cm]{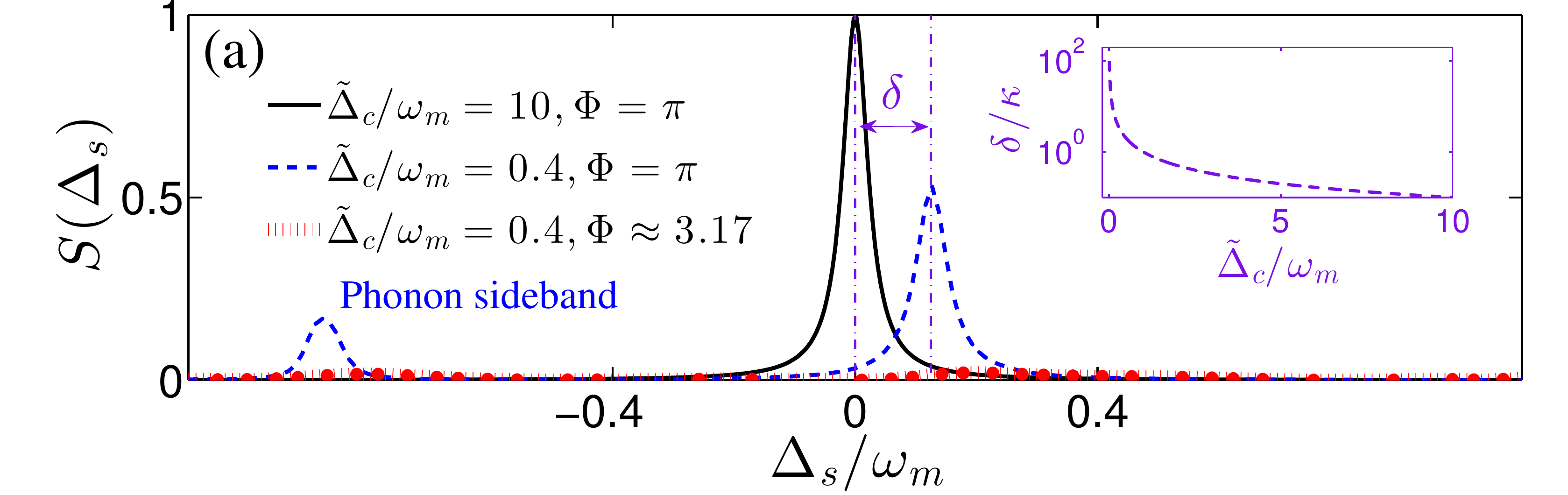}
\includegraphics[width=8cm]{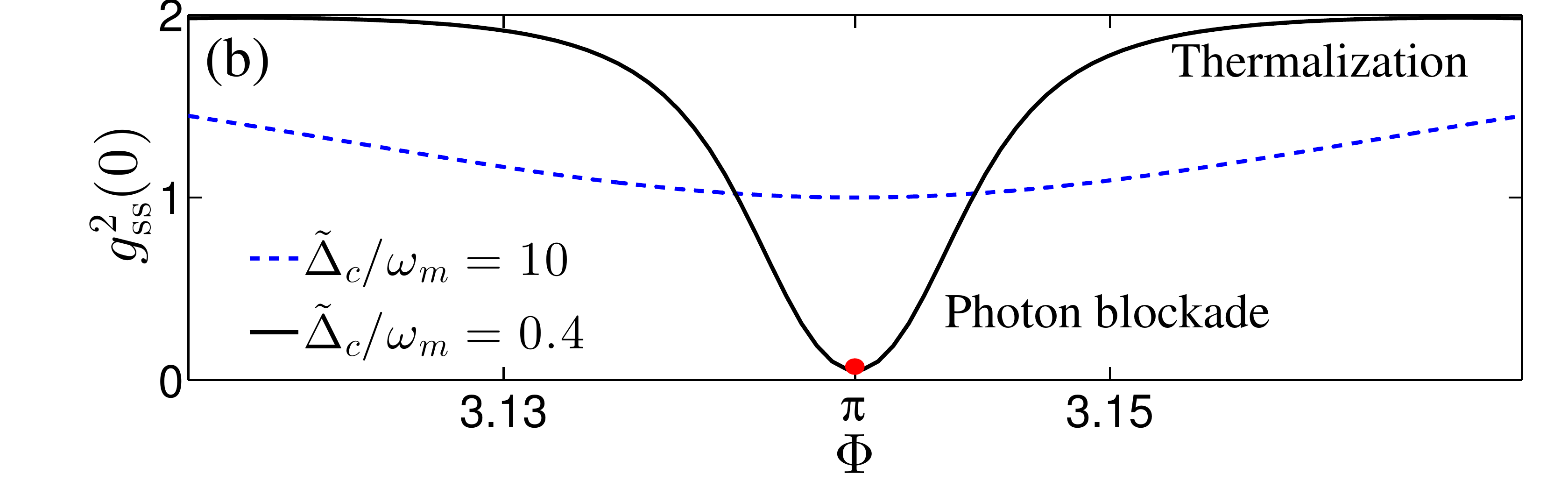}
\includegraphics[width=8cm]{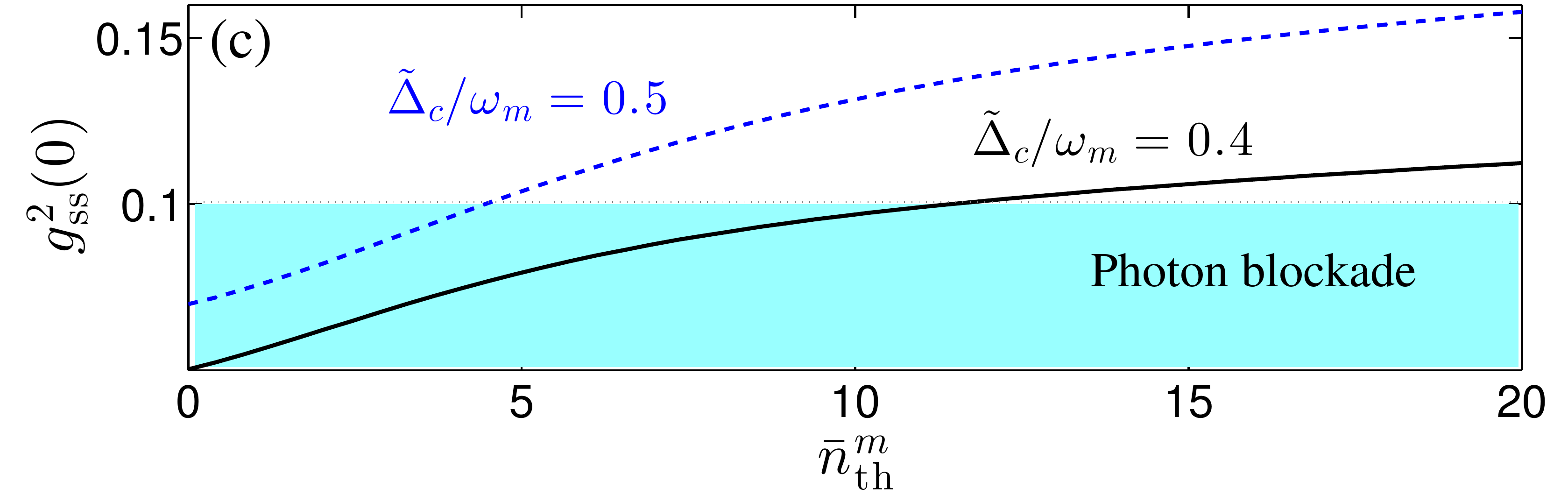}
\includegraphics[width=8cm]{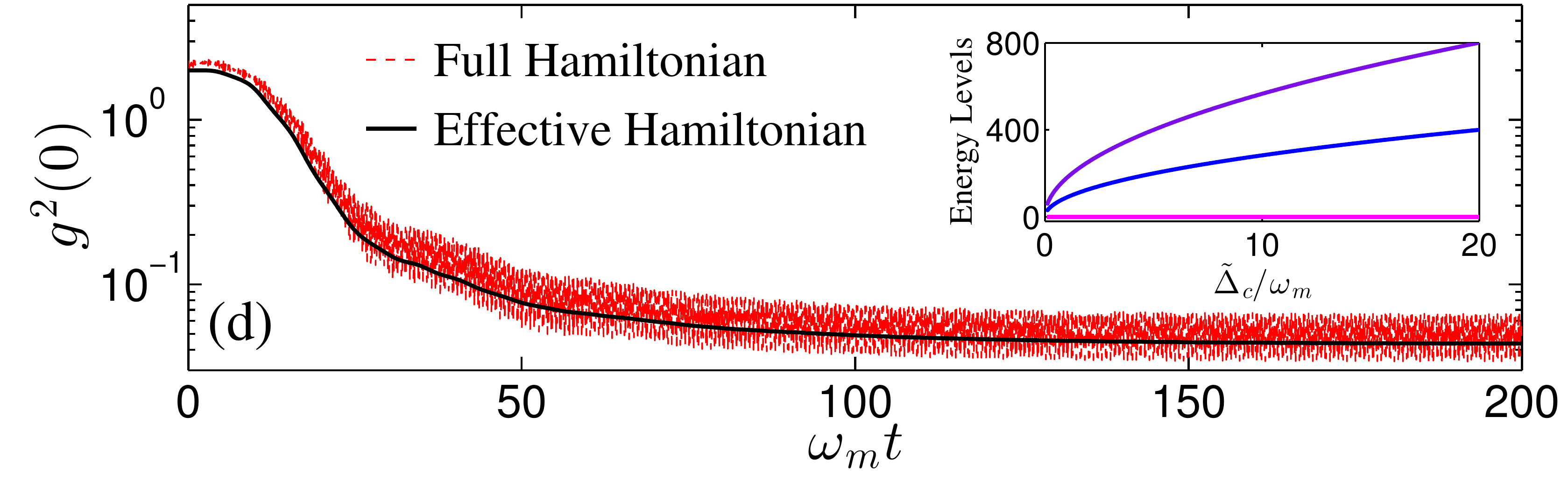}
\caption{(Color online) (a) Cavity excitation spectrum $S(\Delta_s)$ for different $\tilde{\Delta}_c=\Delta_c-2\Lambda$ and $\Phi$. The insert: Shift of the zero-phonon-transition peak $\delta/\kappa$ versus $\tilde{\Delta}_c$. The correlation function $g_{\rm ss}^{2}(0)$ versus (b) $\Phi$ (c) $\bar{n}^{m}_{\rm th}$ for different $\tilde{\Delta}_c$. The shaded area in (c) corresponds to the regime $g_{\rm ss}^{2}(0)<0.1$. (d) $g^{2}(0)$ versus time when $\Delta_s=g^2_s/\omega_m$, and modes $a_s$, $b$ are initially in a thermal state and vacuum state, respectively. The black solid (red dashed) curve is obtained by numerically calculating Eq.\,(\ref{rho}) with $H^d_{\rm OMS}$ ($H_t$). Insert: the three lowest levels of OMS versus $\tilde{\Delta}_c$. The parameters are the same as in Fig.~\ref{fig2} except for $\epsilon_l=10^{-3}\omega_m$, $\Lambda/\omega_m=2000$, and (a,b) $\bar{n}^{m}_{\rm th}=0$, (c,d) $\Phi=\pi$ corresponding to the red point in (b).}
\label{fig4}
\end{figure}

The exact evolution of the system, including the probe field, is also governed by Eq.\,(\ref{rho}), but with the replacement $H\rightarrow H_{t}=H+H_{p}$, where
\begin{align}
\!\!\!\!H_{t}\!\!=\!\!H\!\!+\!\!\epsilon_l[\cosh(r_d)a_s^{\dagger}e^{-i\omega^s_l t}\!\!-\!\sinh(r_d)a_s^{\dagger}e^{i\omega^s_l t\!-\!i\Phi_d}\!\!+\!\!\rm{h.c.}].\label{H_t}
\end{align}
Under the conditions of $\omega^s_l\approx\omega_s$ and $\omega_s\gg \omega_m, g_p, \epsilon_l\sinh(r_d)$, $H_t$ simplifies to $H^d_{\rm OMS}=H_{\rm OMS}+\epsilon_l\cosh(r_d)\{a_s^{\dagger}\exp[-i\omega^s_lt]+\rm{h.c.}\}$ by ignoring the terms oscillating with the high frequencies $2\omega^s_l$, $2\omega_s\pm\omega_m$. In Fig.\,\ref{fig4}(a), we present the excitation spectrum $S(\Delta_s)$ obtained by numerically solving Eq.\,(\ref{rho}) with Hamiltonian $H^d_{\rm OMS}$. It shows that the coupling strength $g_s$ could be obtained by measuring the position of zero-phonon-transition peak $\delta$, since $\delta=g^2_s/\omega_m$  \cite{Rabl2011,Girvin2011,Jieqiao2012}. Moreover, the appearance of phonon sidebands is another signature of the single-photon strong-coupling regime, i.e., $g_s>\kappa$. Fig.\,\ref{fig4}(a) also shows that the spectral information is lost when $\Phi$ deviates too much from its optimal value $\pi$.

Strong radiation-pressure and parametric interactions at the single-photon level provide great potentials for single-photon quantum processes. As an example, we demonstrate the photon blockade, characterized by a vanishing equal-time second-order correlation function in the steady state, $g_{\rm ss}^{2}(0)={\rm Lim}_{t\rightarrow\infty}\langle a^{\dagger}_{s}a^{\dagger}_{s}a_sa_s\rangle(t)/\langle a^{\dagger}_{s}a_s\rangle^2(t)$ and in the transient state, $g^{2}(0)=\langle a^{\dagger}_{s}a^{\dagger}_{s}a_sa_s\rangle(t)/\langle a^{\dagger}_{s}a_s\rangle^2(t)$, when only the $a_s$ mode is weakly driven under the single-photon resonance $\Delta_s=g^2_s/\omega_m$. In Figs.\,\ref{fig4}\,(b) and (c), we plot the dependence of $g^{2}_{\rm ss}(0)$ on $\Phi$ and $\bar{n}^{m}_{\rm th}$, respectively, using $H^d_{\rm OMS}$. They show that the photon blockade occurs in the vicinity of the phase matching $\Phi=\pi$ and for small $\bar{n}^{m}_{\rm th}$. The system is thermalized by the optical noise $N_s$ (or the mechanical noise $\bar{n}^{m}_{\rm th}$) when $\Phi$ deviates too much from $\pi$ (or the temperature of mechanical bath is too high), even in the strong-coupling regime $g_s>\kappa$. Moreover Fig.\,\ref{fig4}(c) also indicates the regime $g_{\rm ss}^{2}(0)<0.1$, corresponding to a strong signature of photon blockade. It shows that photon blockade extends even out to $\bar{n}_{th}^m\sim10$ when $\tilde{\Delta}_c/\omega_m=0.4$. The appearance of photon blockade can be understood qualitatively from the radiation-pressure-induced anharmonicity of the level spacing [see the insert of Fig.\,\ref{fig4}(d)]. Strong anharmonicity makes the probe photons go through the OMS one by one, because the two-photon transition is detuned under the condition of single-photon resonance. The validity of $H^d_{\rm OMS}$ is demonstrated in Fig.\,\ref{fig4}(d), where the evolution of $g^{2}(0)$ corresponding to $H^d_{\rm OMS}$ agrees well with the exact numerical solution using $H_t$ \cite{Johansson2012}. We also note that $g^{2}(0)$ approaches a steady value when $t\approx100/\omega_m$. For $\omega_m=100$ MHz the relaxation time corresponds to 1\,$\mu$s. This requires that the optical (or microwave) driving field has the stable frequency and phase during a time-scale of $\mu$s, which is experimentally feasible with current laser technologies \cite{Young1999,Jiang2011,Fortier2011}.  

Strong radiation-pressure is also useful for cooling a mechanical oscillator.
In sideband cooling experiments, the phase and amplitude noise of the cooling laser induce radiation-pressure fluctuations that ultimately heat the mechanical mode~\cite{Schliesser}, especially in the OMS with a ``soft'' mechanical oscillator~\cite{reviews}. This leads to an excess final occupancy $\bar{n}_{f}$ with a lowest value $\bar{n}_{f}^{\rm min}\propto 1/g_0$ \cite{Rabl2009}. 
Therefore, the enhancement of the radiation-pressure coupling $g_0$ could decrease the practical mechanical-cooling limit by suppressing the influence from the ubiquitous laser noise.

\emph{Conclusions.---} We have presented a method to obtain controllable optomechanical interactions between a squeezed cavity mode and a mechanical mode in an OMS. The squeezed cavity mode is generated by detuned parametric amplification of the original cavity mode, which also interacts with a broadband-squeezed vacuum. We showed that by tuning the intensity or the frequency of the driving field, we can selectively obtain an optomechanical radiation-pressure coupling or a parametric-amplification interaction. Moreover, the effective interaction strengths can potentially be enhanced into the single-photon strong-coupling regime, when originally in the weak-coupling regime. Photon blockade is demonstrated in the vicinity of a phase matching between the broadband squeezed vacuum and the parametric amplification, under which the cavity noise is significantly suppressed. This study provides a promising route for implementing single-photon quantum processes with currently available optomechanical technology.

\emph{Acknowledgements.---} XYL thanks Dr.~Jie-Qiao Liao, Prof.~Yu-xi Liu and Prof.~Peter Rabl for valuable discussions. XYL is supported by NSFC-11374116. YW is supported by NSFC-11375067, and the National Basic Research Program of China (Contract No. 2012CB922103 and No. 2013CB921804). HJ is supported by the NSFC (11274098, 11474087). JZ is supported by the NSFC (61174084, 61134008) and the NBRPC (973 Program) under Grant No. 2014CB921401. F.N. is partially supported by the RIKEN iTHES Project, MURI Center for Dynamic Magneto-Optics, and a Grant-in-Aid for Scientific Research (S).


\begin{thebibliography}{99}
\bibitem{reviews} For recent reviews, see, e.g., M. Aspelmeyer, T.J. Kippenberg, F. Marquardt, Rev. Mod. Phys. \textbf{86}, 1391 (2014); M. Aspelmeyer, P. Meystre and K. Schwab, Physics Today \textbf{65}(7), 2935 (2012); P. Meystre, Ann. Phys. \textbf{525}, 215 (2013); 
F. Marquardt and S.M. Girvin, Physics \textbf{2}, 40 (2009); T.J. Kippenberg and K.J. Vahala, Science \textbf{321}, 1172 (2008).

\bibitem{coolingExp1}A.D. O\textquoteright{}Connell \emph{et al.}, Nature (London) \textbf{464}, 697 (2010).

\bibitem{coolingExp2} J. Chan, T.P.M. Alegre, A.H. Safavi-Naeini, J.T. Hill, A. Krause, S. Gr\"{o}blacher, M. Aspelmeyer, and O. Painter, Nature (London) \textbf{478}, 89 (2011); J.D. Teufel, T. Donner,	D. Li, J.W. Harlow,	M.S. Allman, K. Cicak, A.J. Sirois, J.D. Whittaker, K.W. Lehnert, and R.W. Simmonds, {\it ibid}. \textbf{475}, 359 (2011). 

\bibitem{OITExp1} S. Weis, R. Rivi\'{e}re, S. Del\'{e}glise, E. Gavartin, O. Arcizet, A. Schliesser, T.J. Kippenberg, Science \textbf{330}, 1520 (2010).

\bibitem{OITExp2} A.H. Safavi-Naeini, T.P. Mayer Alegre, J. Chan, M. Eichenfield, M. Winger, Q. Lin, J.T. Hill, D.E. Chang, and O. Painter, Nature (London) \textbf{472}, 69 (2011).

\bibitem{stateconversion1} V. Fiore, Y. Yang, M.C. Kuzyk, R. Barbour, L. Tian, and H. Wang, Phys. Rev. Lett. \textbf{107}, 133601 (2011).

\bibitem{stateconversion2} X. Zhou, F. Hocke, A. Schliesser, A. Marx, H. Huebl, R. Gross, and T.J. Kippenberg, Nat. Phys. \textbf{9}, 179 (2013).

\bibitem{stateconversion3} T.A. Palomaki, J.W. Harlow, J.D. Teufel, R.W. Simmonds, and K.W. Lehnert, Nature (London) \textbf{495}, 210 (2013).

\bibitem{stateconversion4} M. Karuza, C. Biancofiore, M. Bawaj, C. Molinelli, M. Galassi, R. Natali, P. Tombesi, G. Di Giuseppe, and D. Vitali, Phys. Rev. A \textbf{88}, 013804 (2013).

\bibitem{squeezingExp1} D.W. C. Brooks, T. Botter, S. Schreppler, T.P. Purdy, N. Brahms, and D.M. Stamper-Kurn, Nature (London) \textbf{488}, 476 (2012).

\bibitem{squeezingExp2} A.H. Safavi-Naeini, S. Gr\"{o}blacher, J.T. Hill, J. Chan, M. Aspelmeyer, and O. Painter, Nature (London) \textbf{500}, 185 (2013).

\bibitem{squeezingExp3} T.P. Purdy, P.-L. Yu, R.W. Peterson, N.S. Kampel, and C.A. Regal, Phys. Rev. X \textbf{3}, 031012 (2013).

\bibitem{Huang2009} S. Huang and G.S. Agarwal, Phys. Rev. A \textbf{80}, 033807 (2009).

\bibitem{Xiong2012} H. Xiong, L.-G. Si, A.-S. Zheng, X. Yang, and Y. Wu, Phys. Rev. A \textbf{86}, 013815 (2012).

\bibitem{Wang2012} Y.-D. Wang and A.A. Clerk, Phys. Rev. Lett. \textbf{108}, 153603 (2012).

\bibitem{Tian2012} L. Tian, Phys. Rev. Lett. \textbf{108}, 153604 (2012). 

\bibitem{Ma2014} J. Ma, C. You, L.-G. Si, H. Xiong, X. Yang, and Y. Wu, Opt. Lett. \textbf{39}, 4180 (2014).

\bibitem{Jiang2014} C. Jiang, Y. Cui, and K.-D. Zhu, Opt. Express, \textbf{22}, 13773 (2014).

\bibitem{Jing2014} H. Jing, S. K. \"{O}zdemir, X.-Y. L\"{u}, J. Zhang, L. Yang, and F. Nori, Phys. Rev. Lett. \textbf{113}, 053604 (2014).

\bibitem{Bose1997} S. Bose, K. Jacobs, and P.L. Knight, Phys. Rev. A \textbf{56}, 4175 (1997).

\bibitem{Bouwmeester2003} W. Marshall, C. Simon, R. Penrose, and D. Bouwmeester, Phys. Rev. Lett. \textbf{91}, 130401 (2003).

\bibitem{XWXu2013} X.-W. Xu, H. Wang, J. Zhang, and Y. X. Liu, Phys. Rev. A \textbf{88}, 063819 (2013).

\bibitem{Rabl2011} P. Rabl, Phys. Rev. Lett. \textbf{107}, 063601 (2011).

\bibitem{Girvin2011} A. Nunnenkamp, K. B\o rkje, and S.M. Girvin, Phys. Rev. Lett. \textbf{107}, 063602 (2011).

\bibitem{Jieqiao2012} J.Q. Liao, H. K. Cheung, and C.K. Law, Phys. Rev. A \textbf{85}, 025803 (2012).

\bibitem{Ludwig2012} M. Ludwig, A. H. Safavi-Naeini, O. Painter, and F. Marquardt, Phys. Rev. Lett. \textbf{109}, 063601 (2012).

\bibitem{Jieqiao2013} J.Q. Liao and C.K. Law, Phys. Rev. A \textbf{87}, 043809 (2013); J.Q. Liao, K. Jacobs, F. Nori, and R.W. Simmonds, New J. Phys. \textbf{16}, 072001 (2014). 

\bibitem{Yuxi2013} X.-W. Xu, Y.J. Li, and Y.X. Liu, Phys. Rev. A \textbf{87}, 025803 (2013).

\bibitem{Marquardt2013} A. Kronwald, M. Ludwig, and F. Marquardt, Phys. Rev. A \textbf{87}, 013847 (2013).

\bibitem{XYL2013}X.-Y. L\"{u}, W.-M. Zhang, S. Ashhab, Y. Wu, and F. Nori, Sci. Rep. \textbf{3}, 2943 (2013). 

\bibitem{Rabl2012} K. Stannigel, P. Komar, S.J.M. Habraken, S.D. Bennett, M.D. Lukin, P. Zoller, and P. Rabl, Phys. Rev. Lett. \textbf{109}, 013603 (2012).

\bibitem{Lukin2013} P. K\'{o}m\'{a}r, S.D. Bennett, K. Stannigel, S.J.M. Habraken, P. Rabl, P. Zoller, and M.D. Lukin, Phys. Rev. A \textbf{87}, 013839 (2013).

\bibitem{Clerk2013} M.-A. Lemonde, N. Didier, and A.A. Clerk, Phys. Rev. Lett. \textbf{111}, 053602 (2013); K. B\o rkje, A. Nunnenkamp, J.D. Teufel, and S.M. Girvin, Phys. Rev. Lett. \textbf{111}, 053603 (2013); A. Kronwald, F. Marquardt, Phys. Rev. Lett. \textbf{111}, 133601 (2013).

\bibitem{Jieqiao2014} J.Q. Liao and F. Nori, Phys. Rev. A \textbf{90}, 023851 (2014).

\bibitem{Xuereb2012} A. Xuereb, C. Genes, and A. Dantan, Phys. Rev. Lett. \textbf{109}, 223601 (2012).

\bibitem{Xuereb2013} A. Xuereb, C. Genes, and A. Dantan, Phys. Rev. A \textbf{88}, 053803 (2013).

\bibitem{Sillanp2014} T.T. Heikkil\"{a}, F. Massel, J. Tuorila, R. Khan, and M.A. Sillanp\"{a}\"{a}, Phys. Rev. Lett. \textbf{112}, 203603 (2014).

\bibitem{Johansson2014} J.R. Johansson, G. Johansson, F. Nori, Phys. Rev. A \textbf{90}, 053833 (2014).

\bibitem{Nation2014} A.J. Rimberg, M.P. Blencowe, A.D. Armour, and P.D. Nation, New J. Phys. \textbf{16}, 055008 (2014).

\bibitem {Gang2014} G. Li, T. Wang, and H.-S. Song, Phys. Rev. A \textbf{90}, 013827 (2014).

\bibitem{Genes} C. Genes, A. Xuereb, G. Pupillo, and A. Dantan, Phys. Rev. A \textbf{88}, 033855 (2013).

\bibitem{Schnabel2013} S. Ast, M. Mehmet, and R. Schnabel, Opt. Express \textbf{21}, 13572 (2013). 

\bibitem{Devoret2009} A. Kamal, A. Marblestone, and M. Devoret, Phys. Rev. B \textbf{79}, 184301 (2009).

\bibitem{Siddiqi2013} K.W. Murch, S. J. Weber, K.M. Beck, E. Ginossar, and I. Siddiqi, Nature (London) \textbf{499}, 62 (2013).

\bibitem{Gardiner1986} C.W. Gardiner, Phys. Rev. Lett. \textbf{56}, 1917 (1986).

\bibitem{Georgiades1995} N.P. Georgiades, E.S. Polzik, K. Edamatsu, H.J. Kimble, and A.S. Parkins, Phys. Rev. Lett. \textbf{75}, 3426 (1995).

\bibitem{Dayan2004} B. Dayan, A. Peer, A.A. Friesem, and Y. Silberberg, Phys. Rev. Lett. \textbf{93}, 023005 (2004).

\bibitem{Zoller1999} K. J\"{a}hne, C. Genes, K. Hammerer, M. Wallquist, E.S. Polzik, and P. Zoller, Phys. Rev. A \textbf{79}, 063819 (2009). 

\bibitem{GXLi2013} W.-J. Gu, G.-X. Li, and Y.-P. Yang, Phys. Rev. A \textbf{88}, 013835 (2013).

\bibitem{CKLawPRA1995} C. K. Law, Phys. Rev. A \textbf{51}, 2537 (1995).

\bibitem{Agarwal2009} S. Huang and G.S. Agarwal, Phys. Rev. A \textbf{79}, 013821 (2009).

\bibitem{Paternostro} A. Xuereb, M. Barbieri, and M. Paternostro, Phys. Rev. A \textbf{86}, 013809 (2012).

\bibitem{Benlloch} F. Jim\'{e}nez and C. Navarrete-Benlloch, arXiv: 1412.2521.

\bibitem{book2002} H.-P. Breuer and F. Petruccione, {\it The Theory of Open Quantum Systems} (Clarendon Press, Oxford, 2002), Chap. 3.

\bibitem{Johansson2012} J.R. Johansson, P.D. Nation, and F. Nori, Comput. Phys. Commun \textbf{183}, 1760 (2012); \textbf{184}, 1234 (2013).

\bibitem{Young1999} B.C. Young, F.C. Cruz, W.M. Itano and J.C. Bergquist, Phys. Rev. Lett. \textbf{82}, 3799 (1999).
 
\bibitem{Jiang2011} Y.Y. Jiang, A.D. Ludlow, N.D. Lemke, R.W. Fox, J.A. Sherman, L.-S. Ma and C.W. Oates, Nat. Photonics \textbf{5}, 158 (2011).

\bibitem{Fortier2011} T.M. Fortier, M.S. Kirchner, F. Quinlan, J. Taylor, J.C. Bergquist, T. Rosenband, N. Lemke, A. Ludlow, Y. Jiang, C.W. Oates and S.A. Diddams, Nat. Photonics \textbf{5}, 425 (2011).
 
\bibitem{Schliesser} A. Schliesser, R. Rivi\`{e}re, G. Anetsberger, O. Arcizet, and T.J. Kippenberg, Nat. Phys. \textbf{4}, 415 (2008).

\bibitem{Rabl2009} P. Rabl, C. Genes, K. Hammerer, and M. Aspelmeyer, Phys. Rev. A \textbf{80}, 063819 (2009).




\end{thebibliography}
\end{document}